\documentclass[conference]{IEEEtran}

\usepackage[utf8]{inputenc}
\usepackage{amsmath, amssymb}
\usepackage{graphicx}
\usepackage{cite}
\usepackage{hyperref}

\title{LLM-assisted web application functional requirements generation – A case study of four popular LLMs over a Mess Management System}

\author{
    \IEEEauthorblockN{Rashmi Gupta, Aditya K. Gupta, Aarav Jain, Avinash C. Pandey, Atul Gupta}
    \IEEEauthorblockA{
        Department of Computer Science and Engineering \\
        PDPM IIITDM Jabalpur, India \\
        \{rashmig, adityakg, aaravj, avish.p, atul\}@iiitdmj.ac.in
    }
}

\begin{document}

\maketitle
\renewcommand{\thefootnote}{\fnsymbol{footnote}}
\footnotetext[1]{This is the author’s version of a paper accepted at EASE 2025. The final version will appear in the ACM Digital Library.}
\renewcommand{\thefootnote}{\arabic{footnote}}

\begin{abstract}
Like any other discipline, Large Language Models (LLMs) have significantly impacted software engineering by helping developers generate the required artifacts across various phases of software development. This paper presents a case study comparing the performance of popular LLMs—GPT, Claude, Gemini, and DeepSeek — in generating functional specifications that include use cases, business rules, and collaborative workflows for a web application, the Mess Management System. The study evaluated the quality of LLM-generated use cases, business rules, and collaborative workflows in terms of their syntactic and semantic correctness, consistency, non-ambiguity, and completeness compared to the reference specifications against the zero-shot prompted problem statement. Our results suggested that all four LLMs can specify syntactically and semantically correct, mostly non-ambiguous artifacts. Still, they may be inconsistent at times and may differ significantly in the completeness of the generated specification. Claude and Gemini generated all the reference use cases, with Claude achieving the most complete but somewhat redundant use case specifications. Similar results were obtained for specifying workflows. However, all four LLMs struggled to generate relevant Business Rules, with DeepSeek generating the most reference rules but with less completeness. Overall, Claude generated more complete specification artifacts, while Gemini was more precise in the specifications it generated. 
\end{abstract}

\section{Introduction}
Formally specifying software has remained one of the challenging tasks in software engineering. The challenges are apparent: the specifications must be well-defined, unambiguous, complete, consistent, and aligned with the stakeholder needs. Being mostly human-centric, developing more formal specifications poses more technical challenges, whereas natural language issues greatly influence more human-readable specifications. The emergence of large language models has created an opportunity to automate and to enhance RE processes, particularly in drafting Software Requirements Specifications (SRS), validating requirements, prioritizing tasks, and even translating textual requirements into executable software artifacts.

Recent review studies have suggested revolutionizing the software engineering domain by using LLMs across various phases of software development, particularly in code generation, summarization, translation, and quality evaluations~\cite{10449667,10.1145/3695988,zheng2025understanding}. While the capabilities of LLMs are being formally investigated, a growing number of reviews suggest using human-supervised LLM-assisted approaches to be more effective towards optimising software development efforts~\cite{10449667,10.1145/3695988}.

The application of LLMs in RE is also gaining significant traction, with studies exploring their potential for automating key RE tasks such as requirements drafting, validation, prioritization, completeness checking, and transformation into models or code~\cite{10109345,10371698,10628461,luitel2024, sami2024,10628462,10.1007/978-3-031-78386-9_3}. Some preliminary evidence suggests LLMs can translate natural language requirements to a target template or semi-formal language, but rigorous evaluations are needed~\cite{10628852}. LLMs demonstrate promise in improving efficiency and reducing manual effort to the extent of claiming that most software development is to be reduced to requirement engineering and system testing~\cite{10628487,10628462}. However, others see challenges related to accuracy, relevancy, completeness, and domain adaptation~\cite{10109345, Weck_2024}. 

While LLMs can potentially enhance requirement engineering processes~\cite{Arora2024}, their performance in real-world scenarios needs empirical assessments. Factors such as application domains, prompt engineering techniques, and dealing with scale are crucial in determining their usefulness. Moreover, a SWOT analysis of LLM applications in RE highlights the need for better automation and integration into development workflows~\cite{Arora2024}. A recent review of challenges in applying LLMs to RE tasks further emphasizes the necessity of domain-specific assessments and standard evaluation frameworks~\cite{Weck_2024}.

Functional requirements of a software system describe each functionality that a user needs to carry out the intended task to support a specific business process. The description must be clear, consistent, and include all relevant information that can be fed to the code development process and to validate the software. Functional specifications may take more formal or less formal shape, depending on the nature of the software to be developed and the choice of the developers. However, a structural approach is essential to argue for its completeness, consistency, and unambiguous nature. For instance, a comprehensive functional specification can be achieved for domain-specific enterprise software through three key specification artifacts: use cases, workflows, and business rules. \textbf{Use cases} describe interactions between users (actors) and the system in a step-by-step fashion, ensuring all interaction scenarios are explicitly captured~\cite{cockburn2000writing,omg2017uml,kulak2012use,tiwari2019use}. \textbf{Business rules} define the events, conditions, and action logic depicting system behavior, ensuring system compliance with domain requirements and procedures~\cite{herbst1994specification,ross2013business,njonko2014rulecnl}. \textbf{Workflows} capture the interaction sequences involving multiple actors for achieving certain desired objectives~\cite{aalst2002workflow,10.1145/2229156.2229157}. These three artifacts complement each other and strengthen functional requirements specification~\cite{kulak2012use}.

Given the increasing adoption of LLMs in requirements engineering, evaluating their ability to generate and structure these artifacts can provide insights into their effectiveness in automating and enhancing the software requirements specification process. Specifically, we investigate the quality of functional specifications generated by popular LLMs, such as GPT, Gemini, Claude, and DeepSeek, for developing domain-specific enterprise software, to evaluate and highlight their capabilities and limitations in specifying functional requirements in common structural formulations. 

This paper presents a case study using large language models (LLMs) to generate functional specifications for a web application, the Mess Management System. The study assessed the quality of LLM-generated use cases, business rules, and collaborative workflows regarding their syntactic and semantic correctness, consistency, non-ambiguity, and completeness compared to the reference specifications against the zero-shot prompted problem statement. 

Our results suggested that all four LLMs can create syntactically and semantically correct, non-ambiguous specification artifacts, but they may be inconsistent and differ significantly in completeness. The use cases generated by Claude and Gemini included all the reference use cases. On the other hand, GPT generated the least number of reference use cases, but they were the most complete use cases among all LLM-generated reference use cases. Claude's use cases included some redundancy, whereas those from Gemini were precise but with less \textit{Recall}. 

Except for Gemini, all other LLMS generated all the reference workflows. Claude-generated workflows were the most complete, followed by GPT, Gemini, and DeepSeek. All four LLMs struggled to generate relevant Business Rules, with DeepSeek generating the most rules but with less completeness, followed by Gemini, GPT, and Claude. Overall, Claude generated more complete specification artifacts, whereas Gemini was more precise regarding generated specifications. We also encountered some instances, mainly from Claude and DeepSeek, providing additional, relevant details that can enhance the rigor of the functional specification. This suggested the valuable assistance of such LLMs in generating quality specifications. 

The rest of the paper is organized as follows: we present related work in the next section, establishing the study's motivation. We present the research methodology in Section~3, the case study results in Section~4, and insights into how well each model generated template specifications in Section~5, including the validity of our findings. Finally, we present our conclusions in Section 6 and possible future directions.

\section{Related Work}

After demonstrating the successful use of LLMs in assisting coding-level tasks, researchers have been encouraged to use them in other, more involved software engineering phases like RE. Some of the recent studies reported early experience with such tasks. For instance, LLMs have been used to generate Software Requirements Specification (SRS) documents~\cite{10628461}, prioritize requirements in agile frameworks~\cite{sami2024}, transform textual requirements into executable code~\cite{10628487}, and detect incompleteness in requirement specifications~\cite{luitel2024}.

Ronanki et. al. compared requirements generated by five RE experts and ChatGPT over seven different requirements quality attributes and concluded that the ChatGPT-generated requirements were reasonably correct and human-understandable~\cite{10628462}. Lubos et. al. employed Llama-2 to assess the quality of the software requirements as per the ISO 29148 standard. Their study involved software engineers using the LLM’s reviews to identify and improve the specifications' issues~\cite{10371698}. Krishna et. al. used GPT and CodeLlama to generate Software Requirements Specification (SRS) documents and compared them to those produced by entry-level engineers~\cite{10628461}. The authors argued that LLM-generated SRS documents were similar, and in some cases even better. More specifically, GPT helped identify inconsistencies in requirement specifications. However, the authors also found that the other LLM-CodeLlama's assistance was not helpful ~\cite{10628461}. 

While LLMs can potentially help improve RE tasks, their specific performance in real-world scenarios remains to be seen. How domain specificity, prompt engineering techniques, and model scalability influence the successful adoption of LLMs in RE workflow is also unclear. Moreover, a SWOT analysis of LLM applications in RE highlights the need for better automation and integration into software development workflows~\cite{Arora2024}. A recent review of challenges in applying LLMs to RE tasks further emphasizes the necessity of standard evaluation frameworks~\cite{Weck_2024} for better generalizations. 

We conduct a comparative evaluation of multiple LLMs in generating functional requirements for a web application within a case study framework. Specifying functional requirements using structural artifacts should allow us to assess LLMs' abilities to generate syntactically and semantically correct, consistent, unambiguous, and complete specification artifacts in a specific domain. By assessing LLMs in a real-world web application scenario, this research can help understand how well these models align with software engineering best practices and where improvements are needed. The results could inform best practices for selecting and fine-tuning LLMs for RE tasks, ultimately leading to more reliable and efficient requirements generation processes.

\section{Research Methodology}
This section introduces the various elements of our case-study-based research methodology. A single case study design approach focuses on the Mess Management System. This software was designed for our institute to automate the business operations related to two student messes on campus. Following Agile development methodologies, the system was designed using Django for the backend and React.js for the frontend. A detailed functional specification document was already available, which served as the reference specification to evaluate the quality of the functional specification artifacts generated by the four LLMs using suitable prompts. 

\subsection{The Functional Specification}
\begin{figure}[h]
\includegraphics[width=0.45\textwidth]{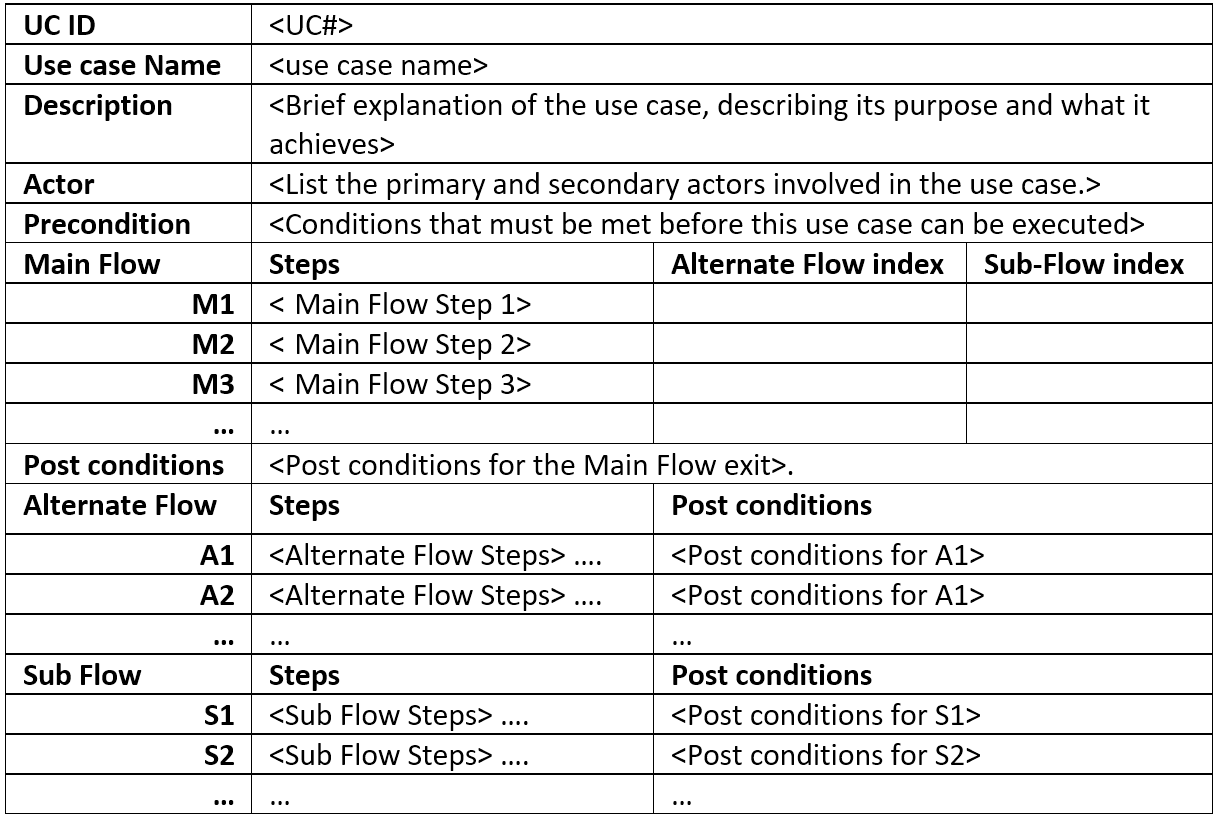}
\caption{The Use case specification template used in the study} \label{uc_template}
\end{figure}

The functional specification consists of three significant key specification artifacts:

\noindent\textbf{Use Cases~\cite{cockburn2000writing,kulak2012use,omg2017uml}} – define the interactions between users (actors) and the system, specifying what functionalities the system provides from a user perspective. Each use case is an asynchronous functionality independently exercised by the actors involved and is documented using the use case template given in Figure~\ref{uc_template}. Use cases can comprehensively describe the functionality of an interactive software system~\cite{tiwari2019use}.

\begin{figure} [h]
\includegraphics[width=0.45\textwidth]{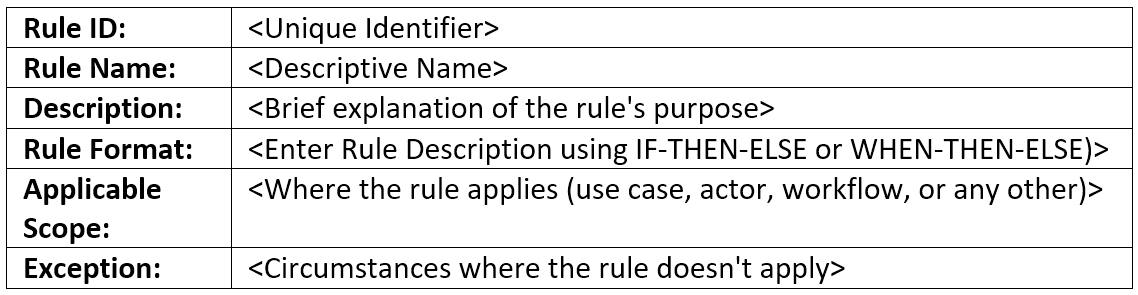}
\caption{The Business rule specification template used in the study} \label{br_template}
\end{figure}

\noindent\textbf{Business Rules~\cite{herbst1994specification,ross2013business,njonko2014rulecnl}} – define the constraints, conditions, and logic governing the system's behavior in different scenarios and use cases. The scope of a business rule may span a use case, an actor, a workflow, or the system. We use a suitable specification format for a business rule as given in Figure~\ref{br_template}.

\begin{figure}[h]
\includegraphics[width=0.45\textwidth]{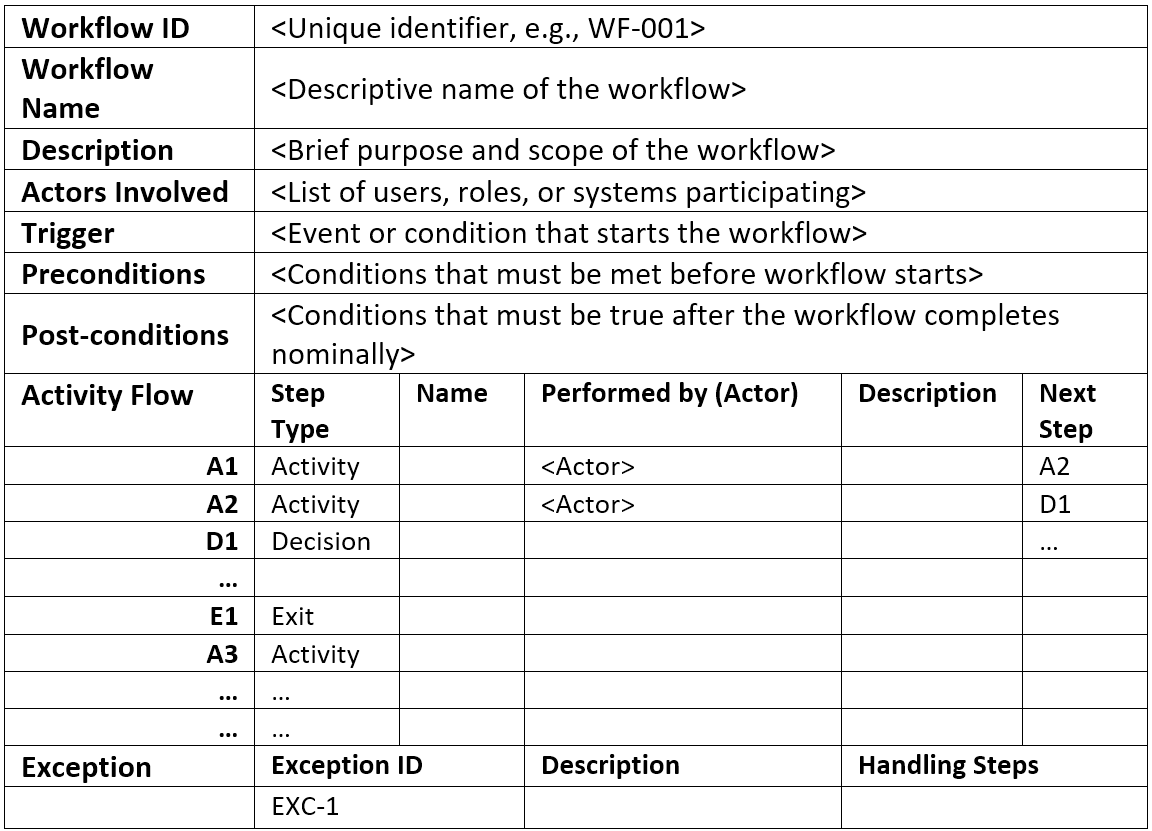}
\caption{The Workflow specification template used in the study} \label{wf_template}
\end{figure}

\noindent\textbf{Workflows~\cite{aalst2002workflow,10.1145/2229156.2229157}} – represent the sequence of activities or steps to achieve specific functionalities involving multiple actors and their specific use cases. Workflows are significant specifications for capturing collaborative work involving multiple actors and span multiple use cases across multiple actors involved. We use a suitable specification format for a workflow as given in Figure~\ref{wf_template}. 

\subsection{The Case Study Problem - Mess Management System}

The Mess Management System supports functionality corresponding to three actors (users) -students, the mess caretaker, and the mess warden - each with specific roles and responsibilities. The system manages mess operations through these three actors, providing functionalities such as user registration/de-registration, announcements, food services, billing, menu reviews and feedback, rebate handling, and general reporting. 

\begin{figure*}
\begin{center}
\includegraphics[width=0.9\textwidth]{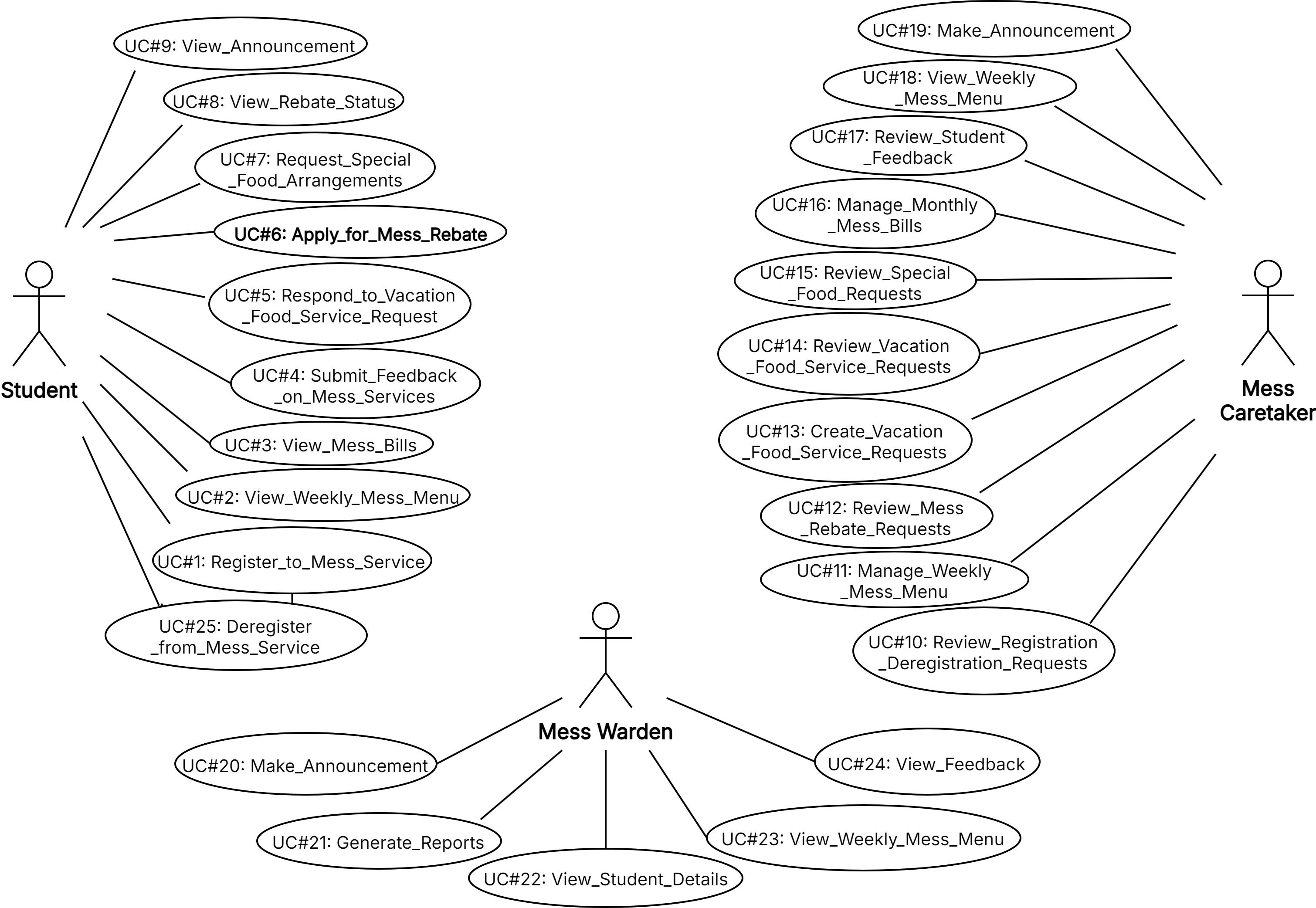}
\caption{The Use case diagram of the Mess Management System} \label{fig1}
\end{center}
\end{figure*}

\begin{figure*}
\begin{center}
\includegraphics[width=0.95\textwidth]{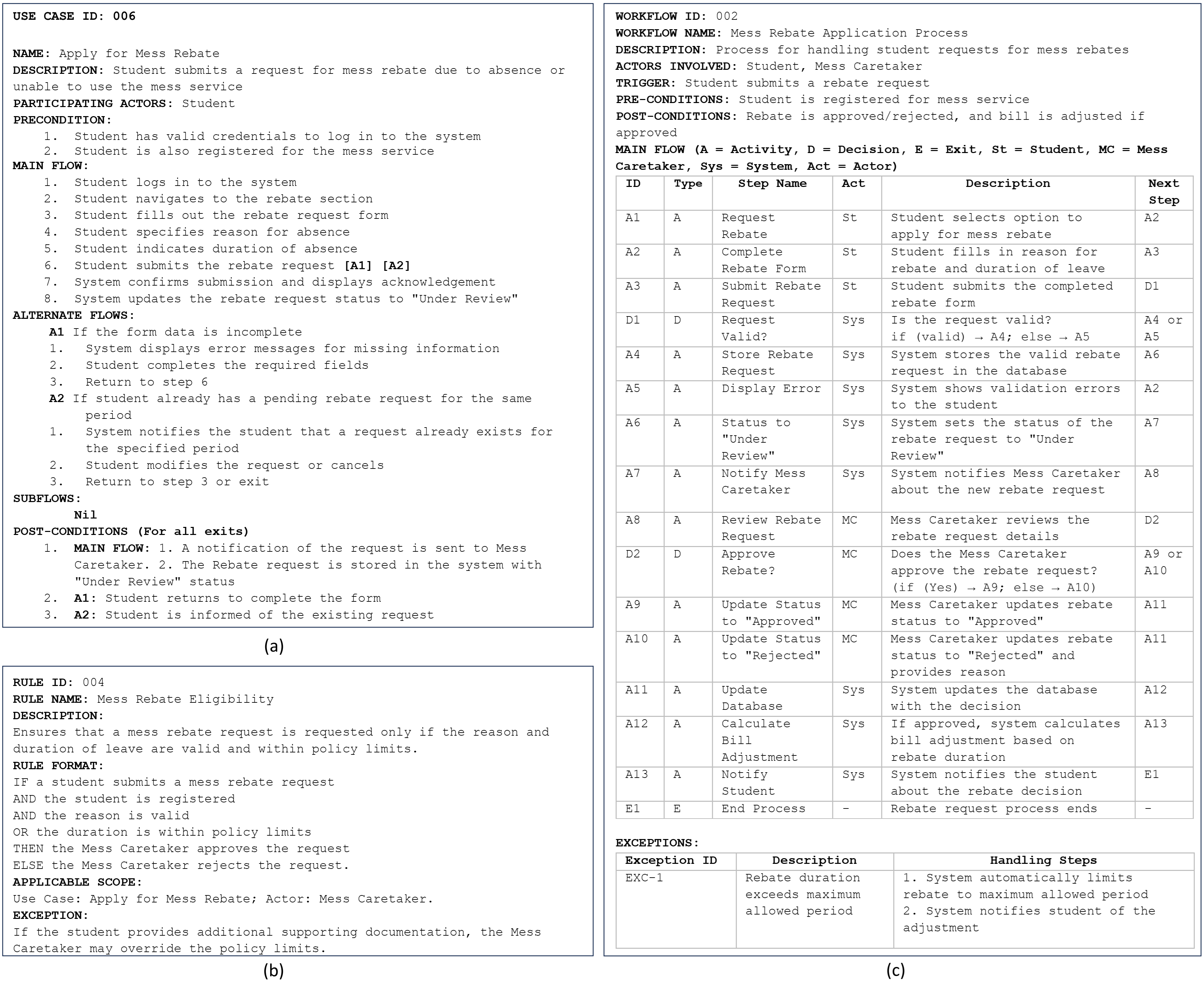}
\caption{Example reference specification artifacts: (a) a use case (b) a business rule, and (c) a workflow} \label{fig2}
\end{center}
\end{figure*}

The use case diagram, depicting actors and their respective use cases, is shown in Figure~\ref{fig1}. Examples of the reference specification artifacts are included in Figure~\ref{fig2} for quick reference. The problem statement, the requirement specification documents consisting of the reference use cases, business rules, and workflows, prompts used to generate the specification artifacts using the four LLMs, the generated specification artifacts, computed data, and the detailed analysis documents are included in~\cite{data_repo}.

\subsection{LLMs used in the study}
This study aims to systematically compare four leading contemporary LLMs—GPT, Gemini, Claude, and DeepSeek—in generating requirements for a web application. These models have been chosen based on their architectural diversity, training methodologies, and relevance to natural language understanding and generation.\\

\noindent\textbf{GPT version 4o (OpenAI~\cite{openai2023gpt4})} – It is perhaps the most popular contemporary LLM, which can understand users’ queries contextually and provide relevant responses. It has been successfully applied in RE tasks such as requirements generation and validation~\cite{10371698,10628461}, making it a strong candidate for evaluation. (knowledge cutoff date - Oct, 2023)\\

\noindent\textbf{Gemini version 2.0 Flash (Google DeepMind~\cite{pichai2023gemini})} – A lightweight and high-speed AI model optimized for low-latency generation. Evaluating Gemini in requirement specification artifact generation could provide insights into how well it understands and handles complex, structured requirements compared to other LLMs. (knowledge cutoff date - Jun, 2024)\\

\noindent\textbf{Claude version 3.7 sonnet (Anthropic~\cite{anthropic2024claude3})} – A model developed with a strong focus on safety, factual correctness, interpretability, and user alignment, Claude emphasis on reducing hallucinations, which makes it an important model to assess in the RE context, where the accuracy and completeness of requirements are crucial. (knowledge cutoff date - Oct, 2024)\\

\noindent\textbf{DeepSeek version V3~\cite{jiang2024deepseek}} – A relatively new entrant in the LLM space, DeepSeek is known for its transparent development approach. Its inclusion in this study should highlight how research-driven alternative models perform against leading industrial models like GPT, Claude, and Gemini in generating domain-specific requirements. (knowledge cutoff date - Jul, 2024)\\

\subsection{Research Questions}
This study investigates the following research questions:
\begin{enumerate}
    \item\textbf{RQ1:} How effective are LLMs in identifying and specifying use cases to be supported by the system while delivering the intended functionalities to its various actors (users)?
    \item\textbf{RQ2:} How effective are LLMs in identifying and specifying business rules, i.e., the additional constraints to be met by the system while delivering the intended functionalities to its various actors?
    \item\textbf{RQ3:} How effective are LLMs in identifying and specifying workflows involving collaborative work among multiple system actors?
\end{enumerate}

\subsection{Criteria for Evaluation}
The effectiveness of the LLM-generated specification artifacts was measured by evaluating the quality of the generated specification. We used the following evaluation criteria to assess the quality of the generated use cases, business rules, and workflow specifications.
\begin{enumerate}
    \item\textbf{Syntactic \& Semantic Correctness:} Measures grammatical accuracy, logical structure, proper use of terminology, and whether statements make sense in context. 
    \item\textbf{Consistency}: Measures internal coherence, ensuring no contradictions exist across different parts of the artifact. Penalizes unnecessary redundancy if it leads to confusion or contradiction. 
    \item\textbf{Non-Ambiguity :} Assesses clarity, ensuring that statements are precise and leave no room for multiple interpretations.  
    \item\textbf{Completeness}: Evaluates whether the generated specification artifact includes all essential components required for a fully functional and well-defined requirement. Here, the completeness of the generated specification artifact is measured against an available reference specification by obtaining measures like \textit{Precision}, \textit{Recall}, and \textit{F1 Score}, which are defined as follows:
    \[
    Precision = \frac{TP}{TP + FP} 
    \]
    \[
    Recall = \frac{TP}{TP + FN}
    \]
    \[
    F1\ Measure = 2*\frac{Precision * Recall}{Precision + Recall}
    \]
    Where, \\
    \textit{True Positive (TP)} - is the number of specification elements included in the generated and reference specification. \\
    \textit{False Positives (FP)} - is the number of additional (unnecessary or incorrect) specification elements included in the generated specification that were not present in the reference specification. \\
    \textit{False Negatives (FN)} - is the number of necessary specification elements missing in the generated specification that were present in the reference specification.
\end{enumerate}

\begin{figure*}
\begin{center}
\includegraphics[width=0.8\textwidth]{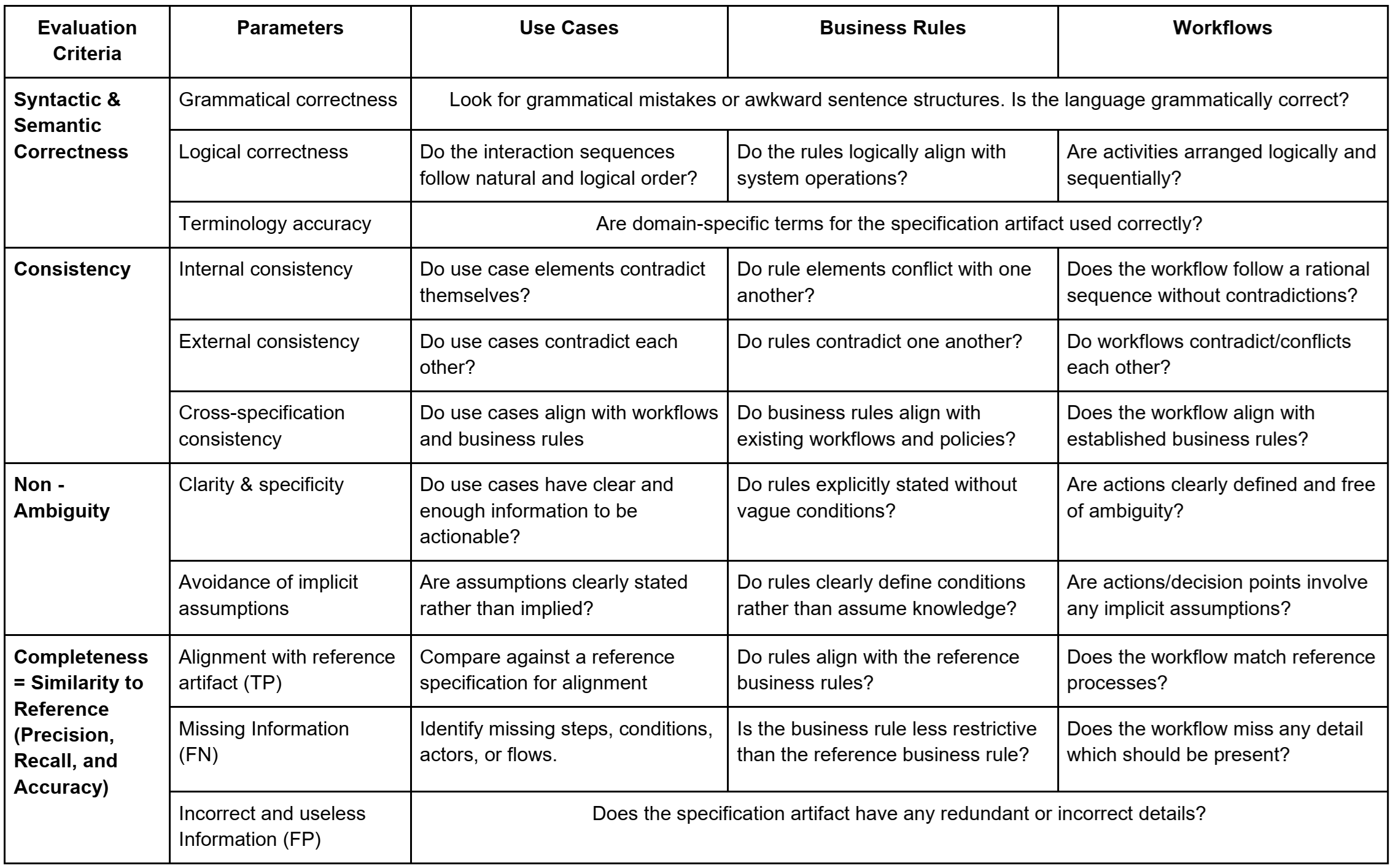}
\caption{The evaluation criteria and assessed parameters of the generated specification artifacts} \label{evaluation-criteria}
\end{center}
\end{figure*}

Figure~\ref{evaluation-criteria} shows how each evaluation criterion was assessed and measured on specific aspects of each functional specification artifact. The first three of the four evaluation criteria are evaluated on a ten-point scale as given in Table~\ref{tab1}. The fourth criterion, i.e., the completeness of the generated specification, is assessed by comparing the generated specification with the reference specification, computing the scores for \textit{TP}, \textit{FP}, and \textit{FN} in the defined manner, and then computing \textit{Precision}, \textit{Recall}, and \textit{F1} scores. 

\begin{table*}
\centering
\caption{Score assigned for syntactic and semantic correctness, consistency, and non-ambiguity criteria}\label{tab1}
\begin{tabular}{|l|l|}
\hline
\textbf{Score} &  \textbf{Interpretation} \\
\hline
0-3 &  Poor: Critical elements missing, incorrect, or unclear \\
4-6 &  Fair: Some issues, partial correctness, but needs improvement \\
7-8 & Good: Mostly correct, but minor improvements needed \\
9-10 & Excellent: Well-structured, complete, and precise \\
\hline
\end{tabular}
\end{table*}


\subsection{Research Procedure}\label{section:resprocess}
The study assesses the abilities of four LLMs—GPT, Claude, Gemini, and DeepSeek—to generate and specify functional specifications using three key specification artifacts: use cases, workflows, and business rules against the problem statement of the Mess Management System. The problem statement describes the functionality to be exercised by the three actors to satisfy the system objectives and purposes. The evaluation process was designed to minimize biases and ensure a fair comparison of the models.

After some initial interactions with the four LLMs, we finalized a zero-shot prompt approach to query each LLM and generate all three functional specification artifacts against the problem statement. The four LLMs were independently prompted to create functional specifications for the problem statement. The models were instructed to structure their output explicitly in the form of the relevant template descriptions, as shown in Figures~\ref{uc_template}, \ref{br_template}, and \ref{wf_template}, respectively.

In pair-evaluation mode, we carried out the scoring process in two phases. In the first phase, two authors used the three generated specification artifacts from the four source LLMs and computed the scores on each of the four evaluation criteria. All the parameters given in Figure~\ref{evaluation-criteria} against each criterion for each reference specification artifact (a use case, a business rule, or a workflow) were assessed on a scale of zero to ten and then averaged to compute the final score against that criterion. In the second phase, two other authors reviewed and finalized scores. The two groups further discussed significant disagreements between scores and were allowed to update their respective scores.  We assessed the inter-rater agreement by computing the Intraclass Correlation Coefficient (ICC)~\cite{ShroutFleiss1979} between the scores obtained in the two phases, which was 0.82, indicating a significant level of agreement between the scores.   

To eliminate evaluator bias, a blind review process was implemented. The generated outputs were anonymized, meaning that the two evaluators were not informed about which LLM had produced which set of artifacts. The responses were randomized and assigned unique identifiers unrelated to the source model. This approach ensured that evaluators focused solely on the quality of the requirements generated rather than any preconceived notions about a specific LLM’s capabilities. 

\section{Results and Analysis}

We undertook the case study as per the procedure given in Section~\ref{section:resprocess}. We collected the data corresponding to the four evaluation criteria against the functional specification artifacts generated by each of the four LLMs. Table~\ref{tab2} shows the number of reference specification artifacts (use cases, business rules, and workflows) generated by each of the four LLMs against the total number of specifications generated. We computed the scores against each criterion for the reference artifacts generated by each of the four LLMs. All the relevant documents for the case study, including zero-shot prompts, reference specifications, generated specifications, and collected data, are available in~\cite{data_repo} for reproducibility and transparency.


In the following sections, we discuss the performance of each of the four LLMs based on the scores computed for the four evaluation criteria against the generated specification artifacts. We used F-1 scores for completeness evaluation. We observed false positives contributed by two major types of discrepancies: irrelevant (or incorrect) and/or redundant details included in the generated specification. We found that the LLMs' outputs were distinctly dominated by these two types of discrepancies, with the former being more serious to deal with. To investigate their individual contributions in obtaining the completeness scores, we computed \textit{Precision (Redundancy)} and \textit{Precision (Incorrectness)} scores separately for both types of discrepancies present in the generated artifacts. 

\begin{table*}
\centering
\caption{Number of specification artifacts generated by the four LLMs}\label{tab2}
\begin{tabular}{|l|l|l|l|}
\hline
& Ref/total              & Ref/total                 &  Ref/total \\
& \# Use cases generated & Business Rules generated  &  Workflows generated \\
& (Ref use cases = 25)   & (Ref business rules  = 10)& (Ref workflows = 4) \\
\hline
GPT     &  17/17  &  3/10   & 4/8 \\
DEEPSEEK  &  19/19  &  5/10   & 4/8 \\
CLAUDE    &  25/30  &  2/25   & 4/5 \\
GEMINI    &  25/28  &  3/16   & 3/6 \\
\hline
\end{tabular}
\end{table*}

\subsection{Use case Identification and Specifications}
\begin{figure}
\includegraphics[width=0.45\textwidth]{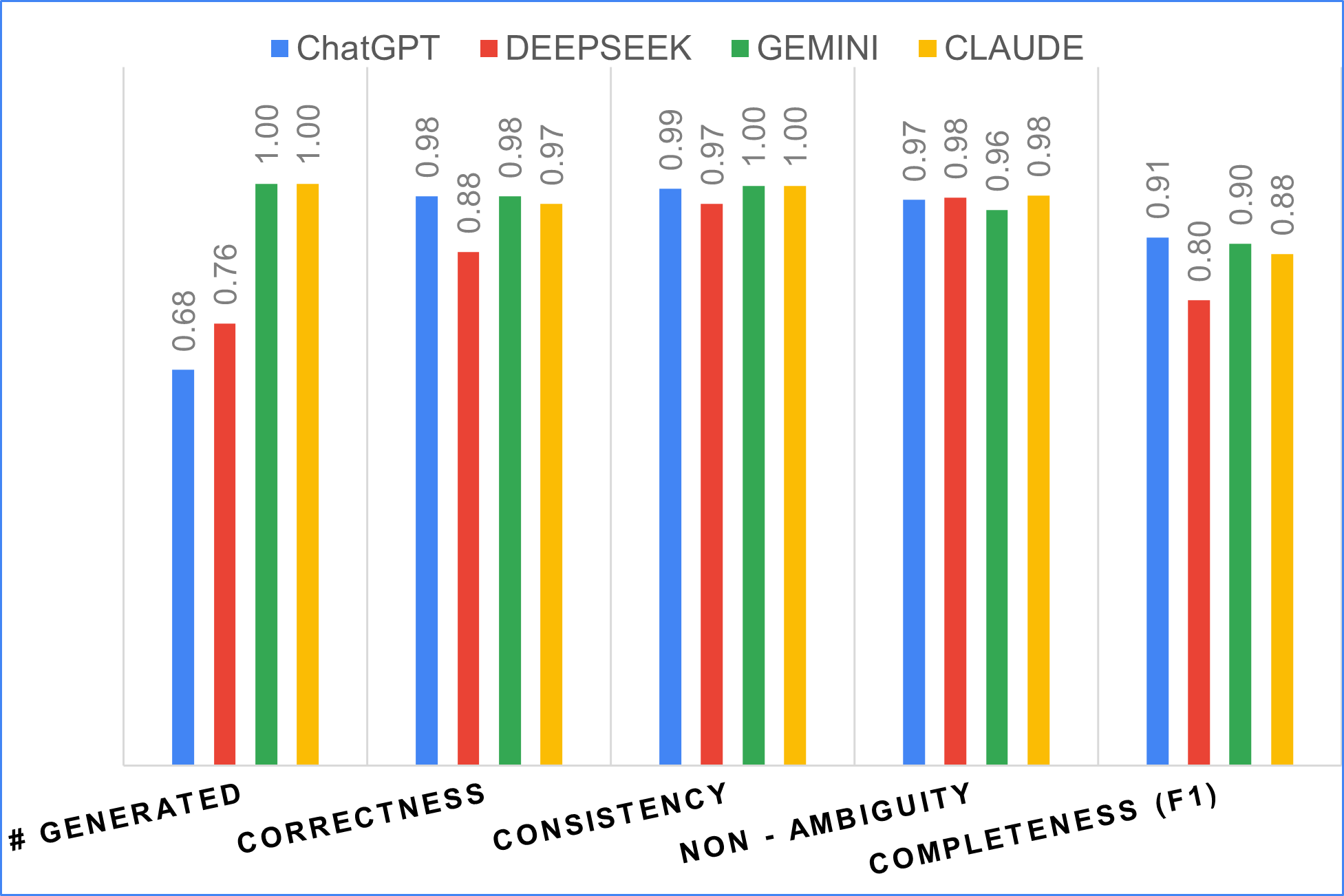}
\caption{Comparison based on Use case specification generated by the four LLMs} \label{UC-comparison-graph}
\end{figure}

Based on the data collected for all twenty-five use cases, the four evaluation criteria and the number of use cases generated by the four LLMs are given in Figure~\ref{UC-comparison-graph}. Gemini and Claude could identify all the relevant use cases, whereas GPT and DeepSeek could identify a subset. All four LLMs' use case specifications had remarkably high syntactic and semantic correctness, consistency, and non-ambiguity values. Except for DeepSeek, the three other LLMs were able to generate significantly complete specifications compared to the reference use case specification.

\begin{figure}
\includegraphics[width=0.45\textwidth]{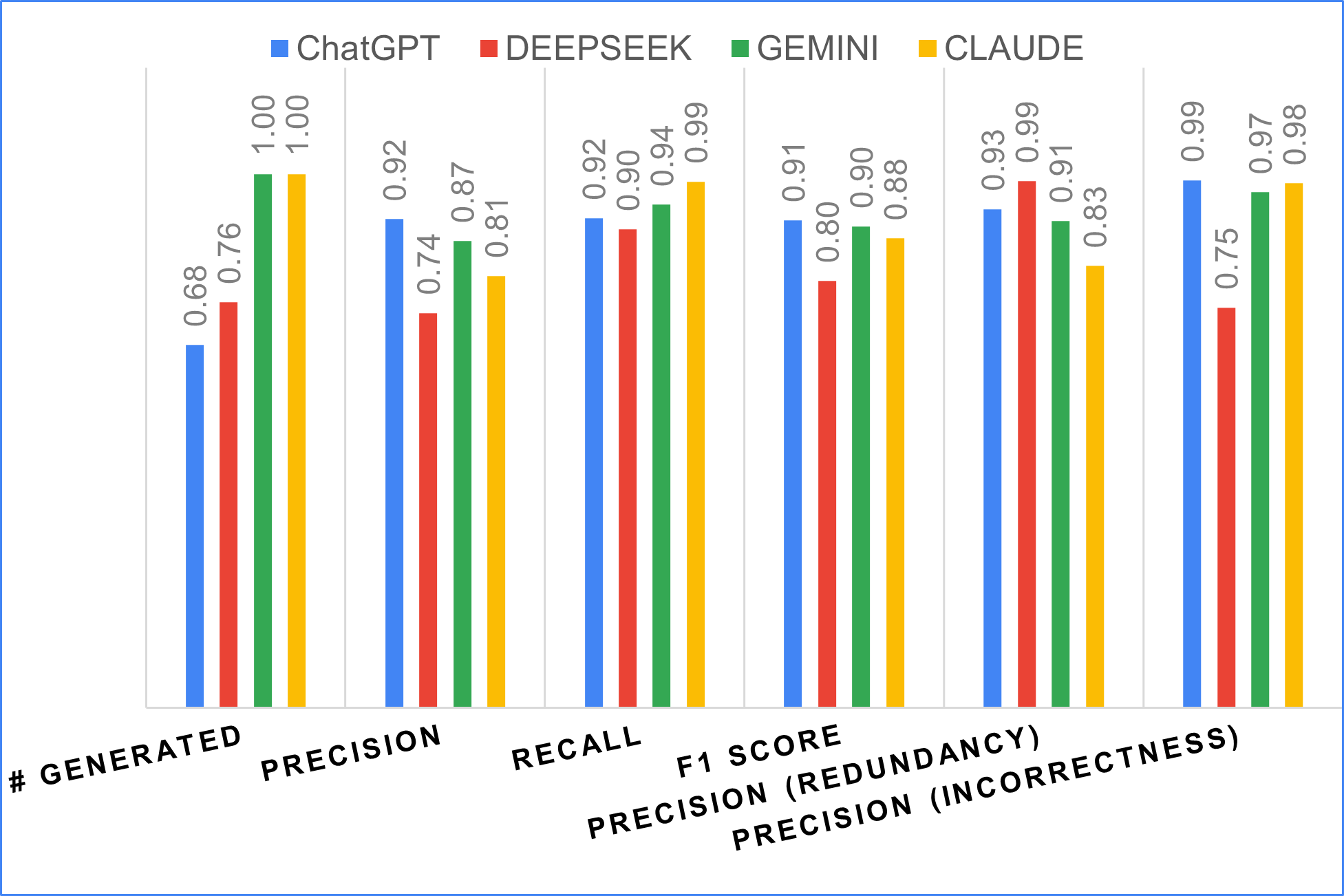}
\caption{Completeness of the Use case specifications generated by the four LLMs} \label{UC_completeness}
\end{figure}


Figure~\ref{UC_completeness} shows the results of a detailed analysis of the generated specifications on the completeness aspect. Though GPT failed to develop some essential use cases, the generated specifications had relatively high \textit{Recall} and \textit{Precision} without missing important details or being redundant. While DeepSeek identified more use cases than GPT, it missed some more significant ones. Although less redundant, DeepSeek scored lowest in various completeness scores because the generated specifications missed some relevant information.

Claude generated all the reference use cases with high \textit{Recall}, but redundant information compromised its \textit{Precision}. On the other hand, Gemini also generated all the reference use cases, but its \textit{Recall} was lower due to missing information. Claude and Gemini had similar \textit{Precision} scores on incorrect information in the generated specifications. For argument's sake, Claude could have outperformed all other LLMs on the resulting F-1 scores if we had underscored the redundant information in the generated use cases.  

In some instances, Claude also provided extra but relevant information that could enhance the rigor of the specification. Examples include suggesting an additional textit{alternate flow} for form validation, reasoning notes about accepting or rejecting requests, and timings for monthly bill calculation in \textit{preconditions}. We did not account for this information when computing redundancy scores.

\subsection{Business Rules Identification and Specification}
\begin{figure}
\includegraphics[width=0.45\textwidth]{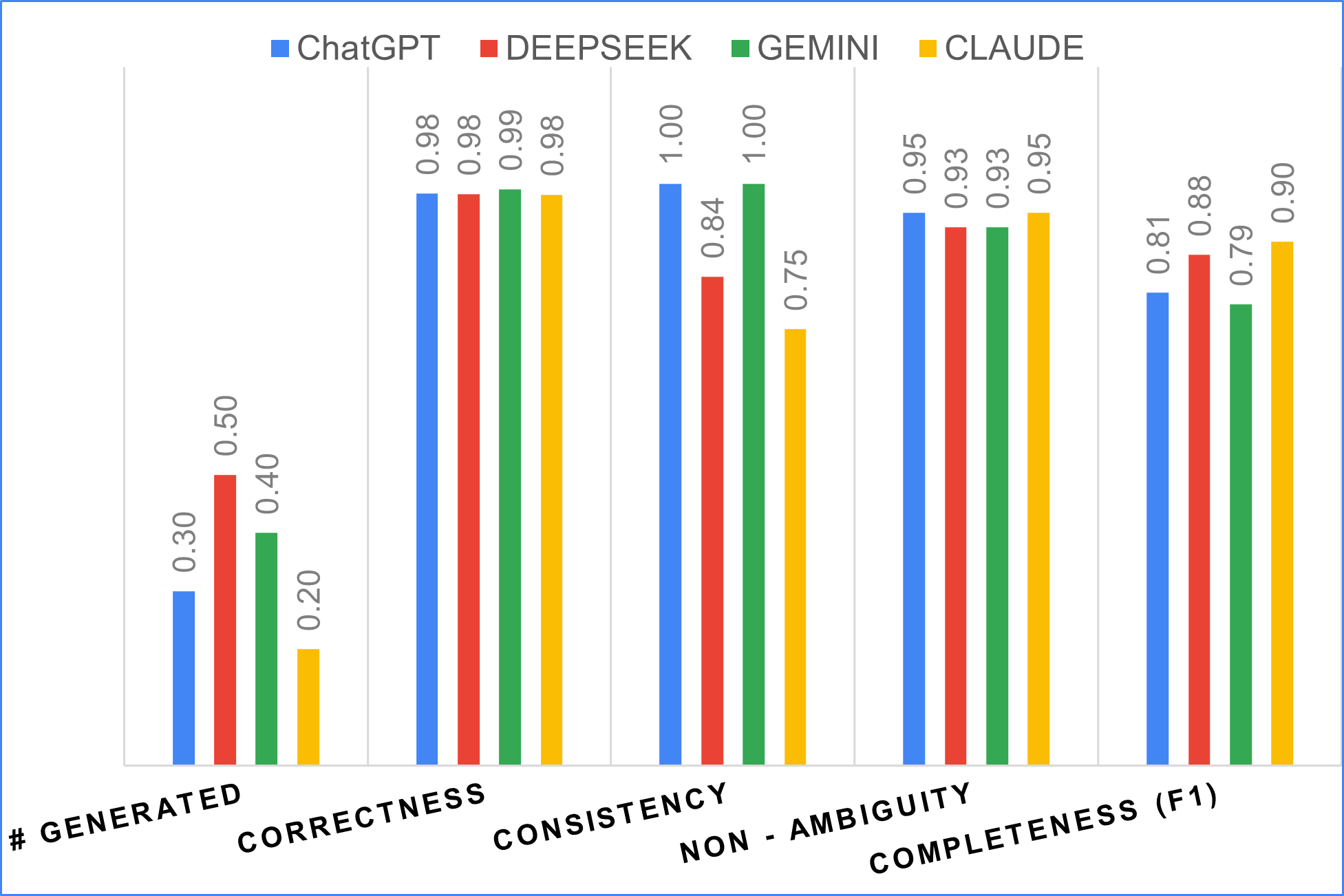}
\caption{Comparison based on Business Rule specifications generated by the four LLMs} \label{BR-comparison-graph}
\end{figure}

The number of reference business rules generated by the four LLMs and scores obtained against the four evaluation criteria are given in Figure~\ref{BR-comparison-graph}. Once again, we found the generated business rules to be high on syntactic and semantic correctness and non-ambiguous. GPT and Gemini produced highly consistent rules, whereas Claude generated rules that included highly redundant information, scoring the lowest on consistency.

All four LLMs had struggled to create relevant business rules, which was a striking difference in the LLM-generated three functional specifications artifacts. DeepSeek generated ten business rules, including five reference business rules, scoring the highest among all LLMs. Claude generated twenty-five, of which we found only two rules relevant to the reference specification. The rest of the rules were either irrelevant or redundant information already included in other generated specifications. We made similar observations about all other LLMs on the generated business rules, which suggested their limitations in creating domain-specific business rules for a given scope.

\begin{figure}
\includegraphics[width=0.45\textwidth]{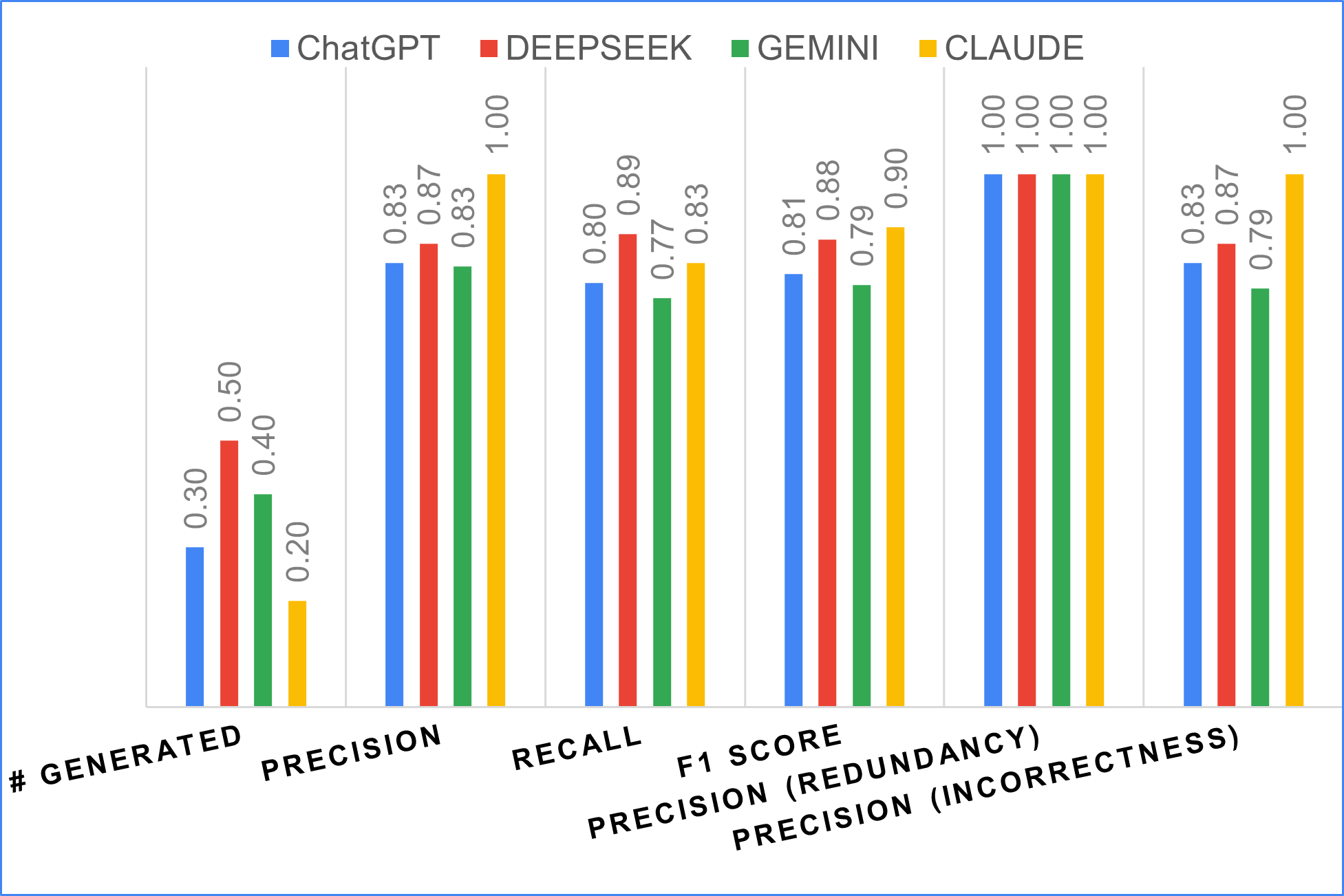}
\caption{Completeness of the Business Rule specifications generated by the four LLMs} \label{BR_completeness}
\end{figure}

Figure~\ref{BR_completeness} shows the results of a detailed completeness analysis of the generated workflow specifications. Although Claud generated only two business rules, it scored highest on the completeness score and achieved the highest precision among the four LLMs. On the other hand, DeepSeek generated the maximum number of reference rules with the highest \textit{Recall} value; its precision was lower than Claud due to some extra irrelevant information in the generated specification. 

\subsection{Workflows Identification and Specification}
\begin{figure}
\includegraphics[width=0.45\textwidth]{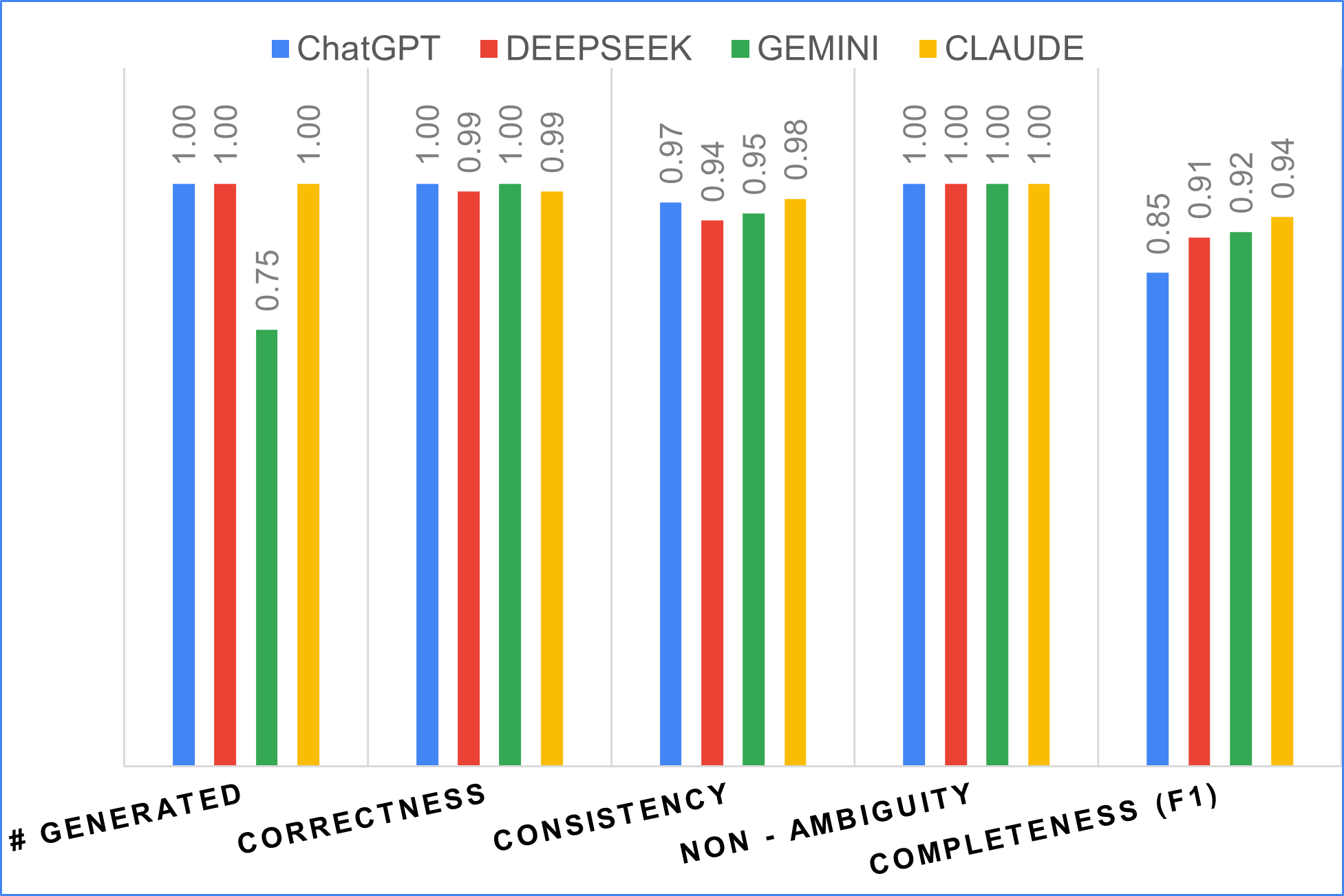}
\caption{Comparison based on Workflow specifications generated by the four LLMs} \label{WF-comparison-graph}
\end{figure}

The number of workflows generated by the four LLMs and scores obtained against the four evaluation criteria are given in Figure~\ref{WF-comparison-graph}. It confirms the syntactic and semantic correctness and unambiguous nature of LLM-generated workflow specifications. Also, these specifications were mainly consistent. Except for Gemini, all other LLMs generated all four reference workflows. 

\begin{figure}
\includegraphics[width=0.45\textwidth]{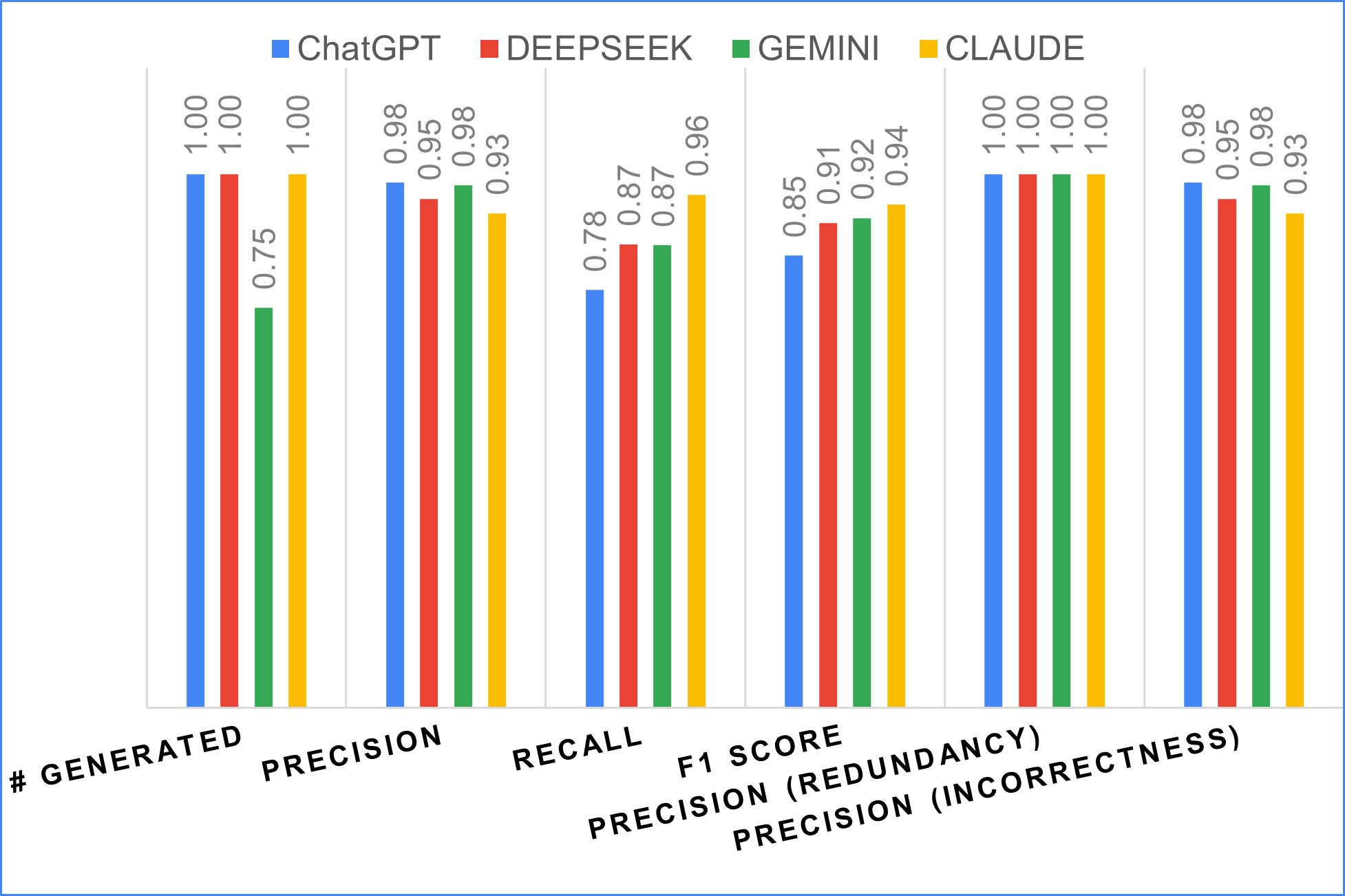}
\caption{Completeness of the Workflow specifications generated by the four LLMs} \label{WF_completeness}
\end{figure}

Figure~\ref{WF_completeness} shows the results of a detailed analysis of the generated workflow specifications on the completeness aspect. Claude outperformed all other LLMs with the highest \textit{Recall}. Again, its low \textit{Precision} was attributed to redundant information. Gemini has missed one workflow, but its precision was highest among generated workflows by all the LLMs. GPT also produced highly precise workflows, but its \textit{Recall} was the lowest among all four LLMs due to missing relevant information.

Again, here, Claude provided extra but useful information that could enhance the rigor of the specification. For example, remind the caretaker of pending requests if no action has been performed. Again, we did not account for such information when computing redundancy scores.

\section{Discussions}
The Mess Management System, although smaller in scope compared to typical enterprise resource planning (ERP) solutions, arguably qualifies as a domain-specific enterprise software application. It supports collaboration among multiple users in a role-based administrative setup, delivering intended functionality such as meal planning, request handling, and billing, within an institutional context through defined organizational business processes and rules.

Here, we acknowledge that LLM outputs are sensitive to the phrasing of the prompt. To mitigate this factor, we refined our prompts through initial testing and ensured the final prompts were task-specific, straightforward, minimal, and unambiguous. The same prompts were given to all the competing LLMs. Still, we admit that a more detailed analysis of the diversity of prompts' phrasing or styling could provide valuable insights on this account.

We found that all the models generated mostly syntactically and semantically accurate responses, with unambiguous specifications, indicating their ability to deal with the most challenging aspects of natural language processing. All models could correctly tell the cross-specification references, indicating that they could connect the generated artifacts logically. However, the generated specifications also had logical inaccuracies and/or redundant information and differed significantly in completeness. Based on our observations in the study, we summarize the nature of the specifications generated by each of the four models as follows.\\

\noindent\textbf{GPT} - generated specifications based on the information explicitly included in the problem statement, and was found to be a little conservative in its approach to providing additional details. So, it was a bit more precise and less redundant in generating specifications. Consequently, its conservative approach resulted in missing out on typical scenarios and was unable to provide more relevant details. \\

\noindent\textbf{Deepseek} - seems to be following a more generalizing approach, relating the problem's details and context to similar situations. Accordingly, it suggested more possible options while generating artifacts, such as business rules. However, some additional information turned out to be irrelevant or incorrect in the specific context, making it less precise for generating artifacts such as use cases and workflows.\\

\noindent\textbf{Gemini} - generated compact and precise information in use cases and workflows, but failed to recognize applicable business rules. It precisely detected the use cases. This compact nature may cause it to miss information, evident from its poorer \textit{Recall}. However, it seemed to follow a conservative approach similar to GPT in providing additional details unless explicitly mentioned.\\

\noindent\textbf{Claude} - provided more detailed and relevant use cases and workflow specifications by enumerating a larger number of alternate business scenarios. It generated all reference use cases and workflows, elaborating each step in detail, which resulted in the highest recall scores. However, this approach also incorporated redundancy, which led to lower precision. At times, it provided additional yet relevant candidate details that could have been included to further enhance the specification. \\

While evaluating the completeness of the LLM-generated specification artifacts against the reference specification, we found that these LLMs generated some extra but relevant information that was not in the reference specification, which did not affect the logical flows, like form validation, or can be a helpful insight like precautions taken by an actor while doing an activity. Accordingly, we ignored them while evaluating the scores. On the other hand, we also encountered additional helpful information that was not in the reference specification. For instance, the use case specification for \textit{make\_announcement} use case generated by Claude included sending an announcement to the target group rather than sending it to the default group. Another example involves DeepSeek suggesting that special food requests be sent at least three days in advance, as per \textit{Business Rule ID 013}. All such instances and models exhibiting high \textit{Precision} and \textit{Recall} suggested significant assistance from LLMs in requirement generation for similar applications.

We followed a blind review process to eliminate bias from the evaluators. Although in this case study, we received valuable and significant insights into the capabilities of these LLMs in generating requirements in a structured specification format, we also acknowledge the limitation of generalizing the observations. The observations are based on a small but real software with a small number of use cases, business rules, and workflows. The workflows are of low-to-moderate complexity. In these circumstances, the observations are evidence of the kind of assistance that can be obtained from different LLMs in requirement specifications. At the same time, our findings confirm the LLM's capabilities for generating high-quality requirement specifications, as observed in other studies~\cite{10628461,10371698,10628462}. 

We summarize the key findings of the study as follows\footnote{The relevant documents for the case study, including problem statement, reference specifications, prompts, collected data, etc., are available in~\cite{data_repo}}:
\begin{enumerate}
    \item All the models could generate syntactic and semantically correct, non-ambiguous functional specification artifacts. However, they differed significantly in consistency (including redundancy) and completeness. 
    \item Overall, Claude generated more complete specification artifacts, whereas Gemini was more precise regarding generated specifications. 
    \item All models struggled to generate relevant business rules.
    \item Some models provided additional, yet relevant, candidate details that could be included to further enhance the specification. 
    \item LLMs can significantly assist in generating and specifying software requirements. Knowing the nature of the LLM to be used in this task can be useful.
\end{enumerate}

\section{Conclusions and Future Directions}

In this work, we empirically assessed the abilities of four contemporary popular LLMs — GPT, Claude, Gemini, and DeepSeek — to specify functional requirements for a web application using three key specification artifacts: use cases, workflows, and business rules. The evaluations were performed on four specification quality criteria - syntactic and semantic correctness, consistency, non-ambiguity, and completeness of the generated specification artifacts.  

The results indicate that LLMs can draft functional requirements to produce relevant, coherent, well-structured use cases and workflows. However, they may struggle to generate relevant business rules to a significant extent. Also, considerable variability was observed in their completeness and, to some extent, in consistency, with Claude generating mostly complete specification artifacts but with some redundancy. In contrast, Gemini's generated specification artifacts were found to be most precise here but with lesser \textit{Recall} values. 

Overall, this study examines the potential of LLMs to assist in generating functional requirements by reducing the manual effort required in the initial phases of software development. However, the generated requirements have to be moderated by human analysts to refine, finalize, and subsequently use these specification artifacts. 

One promising direction is the fine-tuning of LLMs using domain-specific corpora specifically for requirements engineering tasks, as this could enhance their ability to generate more precise and context-aware functional requirements. One can also explore methods such as iterative refinement, few-shot learning, or multi-step prompting, which could help the LLMs understand and specify more complex requirements.

Another potential research direction involves analyzing how well LLM-generated requirements can be used to perform downstream software development tasks, such as software design, generating test cases, and implementing code. Software processes can also be customised to incorporate LLM assistance to optimize development efforts and time.

\bibliographystyle{IEEEtran}
\bibliography{mybib}

\typeout{^^J^^J Job finished: Check for .bbl file now ^^J^^J}

\end{document}